\documentclass[12pt]{article}

\usepackage[english]{babel}
\usepackage{graphicx}

   \usepackage{graphics,color,pstricks,pst-plot,pst-node,pst-coil}
   \usepackage[cp1251]{inputenc}

    \def\lplus{\subset\!\!\!\!\!\!~+}

\begin{document}

\begin{center}
{\large \bf
Particle properties in the early universe from the contraction of the SM gauge group
 }
\end{center}

\begin{center}
 Nikolai A. Gromov \\
 Institute of Physics and Mathematics, Komi Science Centre UrB RAS, \\
Kommunisticheskaya st. 24, Syktyvkar 167 982, Russia \\
gromov@ipm.komisc.ru
\end{center}

\begin{abstract}
The properties of elementary particles and their interactions at different stages of the evolution of the Universe, starting with the Planck energy $ 10 ^{19} $ GeV,  are presented. 
We assume that the Standard Model gauge group becomes simpler as the temperature of the universe increases. The description is based on the hypothesis that the high-energy (high-temperature) limit of the Standard Model is generated by the contraction of the gauge group. An explicit form of the Lagrangian is obtained for each stage of the evolution of the universe and is the basis for describing the properties of elementary particles. These properties change drastically in the infinite temperature limit: all particles lose mass, only massless neutral $ Z $ bosons and $ u $ quarks, as well as neutrinos and photons, survive. Electroweak interactions become long range and are mediated by neutral currents. All quarks are monochromatic.
\end{abstract}



\section{Introduction}

The Standard Model is the modern theory of elementary particles and their interactions. It includes the Electroweak Model, which combines electromagnetic and weak interactions, and Quantum Chromodynamics (QCD), which describes strong interactions. The Standard Model describes the available experimental date quite well, and its adequacy has been convincingly confirmed by the recent discovery of the scalar Higgs boson in the experiments at the Large Hadron Collider. If one wants to investigate the properties and interactions of particles beyond experimentally achieved energies, a possible way is to use the  high-energy (high-temperature) limit of the Standard Model.

 We assume that the gauge group of the Standard Model becomes simpler with increasing energy.
Indeed, the Standard Model is a gauge theory based on a gauge group
 $ SU(3) \times SU(2) \times U(1) $, which is a direct product of simple groups.
Strong interactions of quarks are described by quantumchromodynamics with the $SU(3)$ gauge group and
the characteristic temperature of 0.2 GeV.
The electroweak model is based on the $ SU(2) \times U(1) $ gauge group,
responsible for electroweak interactions with a characteristic temperature of $ 100 $ GeV,
while the $ U(1) $ group is associated with long-range electromagnetic interactions.
Due to the zero mass of the photon  its characteristic temperature extends to the "infinite"  Planck energy $ 10^{19} $ GeV.
It follows from this observation that the gauge group of the theory of elementary particles becomes simpler
with increasing temperature of the Universe. We assume that with further increase in the temperature,
simplification of the gauge group of the Standard Model is described by its contraction.

The operation of contraction (or limiting transition) of groups \cite{IW-53}, which, in particular, transforms a
simple group into a nonsemisimple one, is well known in physics. The notion of contraction was extended to
algebraic structures, such as quantum groups and supergroups, and to fundamental representations of
unitary groups \cite{Gr-12}. For a symmetric physical system, contraction of its symmetry group means a transition
to the limiting state of the system. In the case of a complex physical system, which is the Standard Model,
the study of its limiting states at any given limiting values of physical parameters allows a better insight into
the behavior of the system as a whole. We will discuss the modified Standard Model with the contracted
gauge group at the level of classical gauge fields.

In the broad sense of the word, deformation is the reverse of contraction. Nontrivial deformation of an
algebraic structure means, generally speaking, its nonobvious generalization. A prominent recent example is
quantum groups \cite{FRT}, i.e., such generalizations of Hopf algebras that are simultaneously noncommutative and
noncocommutative, while previously Hopf algebras with only one of these properties were known. However,
if a mathematical structure is first contracted, the initial structure can be reconstructed using deformation
in the narrow sense, performed in the direction opposite to that of contraction.

We use this technique to reconstruct the evolution of elementary particles in the early Universe relying on
the currently achieved knowledge. To this end, we consider the behavior of the Standard Model in the
limit of the "infinite" temperature, which, according to our hypothesis, is generated by contraction of the $SU(2)$ and   $SU(3)$
gauge groups \cite{Gr-2016}.
Similar "infinitely" high temperatures could exist in the early Universe in the first instants after the Big Bang
\cite{GoR-11,L-1990}.
 It turns out that the gauge group contraction results in the Standard Model Lagrangian breaking down into a
few terms with different powers of the zero-tending contraction parameter $ \epsilon \rightarrow 0 $. Since the average
energy (temperature) in the hot Universe is related to its age, then moving forward in time, i.e., in the direction
opposite to the high-temperature contraction, we come to the conclusion that after the birth of the Universe
the elementary particles and their interactions pass through a number of stages in their evolution
from the limiting state with the "infinite" temperature to the state described by the Standard Model. These
stages of quark-gluon plasma formation and reconstruction of electroweak and color symmetries differ
by the powers of the contraction parameter and consequently by the time of their origin. Based on the Standard
Model contraction, we can classify these stages according to the "earlier-later" principle but cannot
find the time elapsed after the birth of the Universe. To establish the absolute time, we use additional
assumptions.
The paper is an expanded and supplemented version of the report at the conference dedicated to the 110th anniversary of the birth of N.N. Bogolyubov \cite{Gr-2020}.


\section{Electroweak Model }


The elementary particles of the  Standard Model are as follows:
scalar  gauge bosons ($
\mbox{photon}\; \gamma, \;\;
\mbox{charged } \; W^{\pm} \;\mbox{ and neutral }  \; Z^{0} \;\;
 \mbox{weak bosons}, \;
\mbox{gluons}\; A^k,\; k=1,\ldots,8),
$
a special particle
 $
\chi \; \mbox{(scalar Higgs boson)},
$
as well as vector particles, namely three generations of leptons
$
\left(
\begin{array}{c}
	\nu_{e} \\
	e
\end{array} \right), \;
$
$
\left(
\begin{array}{c}
	\nu_{\mu} \\
	\mu
\end{array} \right), \;
$
$
\left(
\begin{array}{c}
	\nu_{\tau} \\
	\tau  
\end{array} \right)  \;
$
and three generations of quarks
$
 \left(
\begin{array}{c}
	u \\
	d 
\end{array} \right), \;
$
$
\left(
\begin{array}{c}
	c \\
	s 
\end{array} \right), \;
$
$
\left(
\begin{array}{c}
	t \\ 
	b 
\end{array} \right).
$

We will briefly describe the main features of the Electroweak Model
 according to \cite{R-99}. The  Lagrangian of the model is given by the sum
of the boson, lepton, and quark Lagrangians
$
L=L_B + L_L + L_Q.
$
It  is taken to be invariant under
the action of the gauge group $SU(2)\times U(1)$
in the space  $\mathbf{C}_2 $:
$$
 SU(2):\; \vec{z}\,'=G\vec{z},
$$
$$
\left(\begin{array}{c}
z'_1 \\
z'_2
\end{array} \right)
=\left(\begin{array}{cc}
	\alpha & \beta   \\
-\bar{\beta}	 & \bar{\alpha}
\end{array} \right)
\left(\begin{array}{c}
z_1 \\
z_2
\end{array} \right), \quad
|\alpha|^2+|\beta|^2=1,
$$
\begin{equation}
U(1): \; \vec{z}\,'=e^{i\omega/2}\vec{z}=e^{i\omega Y}\vec{z},\quad \omega\in \mathbf{R}.
\label{eq1qq}
\end{equation}

 The  $U(1)$ group generator $Y$ is proportional to unit matrix $Y=\frac{1}{2}{\bf 1}$.
Generators of $SU(2)$ group
$$
  T_1= \frac{1}{2}\left(\begin{array}{cc}
	0 & 1 \\
	1 & 0
\end{array} \right)= \frac{1}{2}\tau_1, \quad
T_2= \frac{1}{2}\left(\begin{array}{cc}
	0 & -i \\
	i & 0
\end{array} \right)= \frac{1}{2}\tau_2, \quad
$$
\begin{equation}
T_3= \frac{1}{2}\left(\begin{array}{cc}
	1 & 0 \\
	0 & -1
\end{array} \right)= \frac{1}{2}\tau_3,
\label{g7}
\end{equation}
where $\tau_k, \;  k=1,2,3$ are Pauli matrices are subject to commutation relations
\begin{equation}
[T_1,T_2]=iT_3, \quad [T_3,T_1]=iT_2,\quad
 [T_2,T_3]=iT_1
\label{g8}
\end{equation}
and form Lie algebra $su(2).$

The boson sector $L_B=L_A + L_{\phi}$ includes the Lagrangian of the gauge fields
$$
L_A=
 -\frac{1}{4}[(F_{\mu\nu}^1)^2 +(F_{\mu\nu}^2)^2+(F_{\mu\nu}^3)^2] -\frac{1}{4}(B_{\mu\nu})^2,
$$
$$
 F^1_{\mu\nu}=\partial_{\mu}A^1_{\nu}-\partial_{\nu}A^1_{\mu}+ gA^2_{\mu} A^3_{\nu}, \quad
  F^2_{\mu\nu}=\partial_{\mu}A^2_{\nu}-\partial_{\nu}A^2_{\mu}+ gA^3_{\mu}A^1_{\nu},
$$
\begin{equation}
  F^3_{\mu\nu}=\partial_{\mu}A^3_{\nu}-\partial_{\nu}A^3_{\mu}+ gA^1_{\mu}A^2_{\nu}, \quad
  B_{\mu\nu}=\partial_{\mu}B_{\nu}-\partial_{\nu}B_{\mu}
\label{eq2}
\end{equation}
and the  Lagrangian of the matter fields
\begin{equation}
  L_{\phi}= \frac{1}{2}(D_\mu \phi)^{\dagger}D_\mu \phi -
  \frac{\lambda }{4}\left(\phi^{\dagger}\phi- v^2\right)^2, \quad
\phi= \left(
\begin{array}{c}
	\phi_1 \\
	\phi_2
\end{array} \right) \in \mathbf{C}_2.
\label{eq3}
\end{equation}
The covariant derivatives are
 \begin{equation}
D_\mu\phi=\partial_\mu\phi -ig\left(\sum_{k=1}^{3}T_kA^k_\mu \right)\phi-ig'YB_\mu\phi,
\label{eq4}
\end{equation}
where   $T_k=\frac{1}{2}\tau_k,\; k=1,2,3$ are the   $SU(2)$ generators,
 $Y=\frac{1}{2}{\bf 1}$ is  the  $U(1)$ generator, $\tau_{k}$ are the Pauli matrices, and $g$ and $g'$ are the charges.
The gauge fields
 \begin{equation}
 A_\mu (x)=-i\sum_{k=1}^{3}T_kA^k_\mu (x),\quad B_\mu (x)=-iB_\mu (x)
\label{gen}
\end{equation}
take their values in Lie algebras $su(2),$  $u(1)$ respectively,  and their 
field strength  tensors  defined as follows
$$
F_{\mu\nu}(x)={\cal F}_{\mu\nu}(x)+[A_\mu(x),A_\nu(x)],\quad
{\cal F}^k_{\mu\nu}(x)= \partial_{\mu}A^k_{\nu}(x) -\partial_{\nu}A^k_{\mu}(x),
$$
 \begin{equation}
B_{\mu\nu}(x)=\partial_{\mu}B_{\nu}(x) -\partial_{\nu}B_{\mu}(x).
\label{gen-5}
\end{equation}

The new gauge fields
$$
{ Z_\mu(x) =\frac{1}{\sqrt{g^2+g'^2}}\left( gA_\mu^3(x)-g'B_\mu(x) \right)}, \quad
 { A_\mu(x) =\frac{1}{\sqrt{g^2+g'^2}}\left( g'A_\mu^3(x)+gB_\mu(x) \right)},
$$
\begin{equation}
{W_\mu^{\pm}(x)=\frac{1}{\sqrt{2}}\left(A_\mu^1(x)\mp iA_\mu^2(x)  \right)}
  \label{eq5-1}
\end{equation}
are introduced instead of (\ref{gen}).

To generate   mass  for the vector bosons
the special mechanism of spontaneous symmetry breaking
is used.
 One of the $L_B$ ground states
 \begin{equation}
  \phi^{vac}=\left(\begin{array}{c}
	0  \\
	v
\end{array} \right), \quad  A_\mu^k=B_\mu=0
\label{gen-6}
\end{equation}
is taken as a vacuum state of the model, and small field excitations $v+\chi(x) $
with respect to this vacuum  are regarded.

After  spontaneous symmetry breaking
the boson Lagrangian (\ref{eq2}), (\ref{eq3}) can be represented in the form
$$
 L_B=L_B^{(2)} + L_B^{int}=
$$
$$
=\frac{1}{2}\left(\partial_\mu\chi \right)^2 -\frac{1}{2}m_{\chi}^2\chi^2
+\frac{1}{2}m_Z^2Z_\mu Z_\mu - {\frac{1}{4}{\cal Z}_{\mu\nu}{\cal Z}_{\mu\nu}}-
$$
\begin{equation}
-{\frac{1}{4}{\cal F}_{\mu\nu}{\cal F}_{\mu\nu}} 
+m_W^2W_\mu^{+}W_\mu^{-} -{\frac{1}{2}{\cal W}_{\mu\nu}^{+}{\cal W}_{\mu\nu}^{-} } +L_B^{int},
\label{eq5-2}
\end{equation}
where
$
{\cal F}_{\mu\nu}(x)= \partial_{\mu}A_{\nu}(x) -\partial_{\nu}A_{\mu}(x), \;
{\cal Z}_{\mu\nu}(x)= \partial_{\mu}Z_{\nu}(x) -\partial_{\nu}Z_{\mu}(x), \;
{\cal W^{\pm}}_{\mu\nu}(x)= \partial_{\mu}W^{\pm}_{\nu}(x) -\partial_{\nu}W^{\pm}_{\mu}(x).
$
As usual the  second order terms describe the boson particles content of the model and higher order terms $L_B^{int}$  are regarded as their  interactions.
So Lagrangian (\ref{eq5-2}) includes  charged $W$ bosons  with identical mass  $m_W=\frac{1}{2}gv$,
massless photon  $A_\mu, $
 neutral $Z$ boson  with the mass  $m_Z=\frac{v}{2}\sqrt{g^2+g'^2}$
and  scalar Higgs boson  $ \chi,\; m_{\chi}=\sqrt{2\lambda}v$.
All these particles are experimentally detected and have the masses:
$m_W=80\; GeV$, $m_Z=91\; GeV$, $m_{\chi}=125\; GeV$.

The interaction Lagrangian  $L_B^{int}$ looks as follows

$$
L_B^{int}=\frac{g m_z}{2\cos \theta_W} \left(Z_{\mu}\right)^2 \chi -\lambda v \chi^3
+\frac{g^2 }{8\cos^2\theta_W} \left(Z_{\mu}\right)^2 \chi^2 - \frac{\lambda}{4} \chi^4 -
$$
$$
 -{\frac{1}{2}{\cal W}_{\mu\nu}^{+}{\cal W}_{\mu\nu}^{-} + m_W^2W_\mu^{+}W_\mu^{-} } 
$$
$$
-2ig\left(W_\mu^{+}W_\nu^{-} - W_\mu^{-}W_\nu^{+}\right)
\left({\cal F}_{\mu\nu}\sin \theta_W  \right. 
\left.+{\cal Z}_{\mu\nu}\cos \theta_W\right) -
$$
$$
-\frac{i}{2}e \left[A_{\mu}\left({\cal W}_{\mu\nu}^{+}W_\nu^{-} - {\cal W}_{\mu\nu}^{-}W_\nu^{+}\right)  \right.
\left. -A_{\nu}\left({\cal W}_{\mu\nu}^{+}W_\mu^{-} - {\cal W}_{\mu\nu}^{-}W_\mu^{+}\right) \right] +
$$
$$
+gW_\mu^{+}W_\mu^{-}\chi 
-\frac{i}{2}g\cos \theta_W  \left[Z_{\mu}\left({\cal W}_{\mu\nu}^{+}W_\nu^{-} - {\cal W}_{\mu\nu}^{-}W_\nu^{+}\right) - \right.
$$
$$
\left. -Z_{\nu}\left({\cal W}_{\mu\nu}^{+}W_\mu^{-} - {\cal W}_{\mu\nu}^{-}W_\mu^{+}\right) \right] 
+ \frac{g^2}{4}\left(W_\mu^{+}W_\nu^{-} - W_\mu^{-}W_\nu^{+}\right)^2 +
$$
$$
+\frac{g^2}{4}W_\mu^{+}W_\nu^{-}\chi^2
-\frac{e^2}{4} \left\{
\left[\left(W_{\mu}^{+}\right)^2 + \left(W_{\mu}^{-}\right)^2\right](A_{\nu})^2  \right. -
$$
$$
-2\left(W_\mu^{+}W_\nu^{+} + W_\mu^{-}W_\nu^{-} \right)A_{\mu}A_{\nu} +
\left. \left[\left(W_{\nu}^{+}\right)^2 + \left(W_{\nu}^{-}\right)^2\right](A_{\mu})^2
\right\} -
$$
$$
-\frac{g^2}{4}\cos\theta_W \left\{
\left[\left(W_{\mu}^{+}\right)^2
+ \left(W_{\mu}^{-}\right)^2\right](Z_{\nu})^2 -  \right.
$$
$$
 -2\left(W_\mu^{+}W_\nu^{+} + W_\mu^{-}W_\nu^{-} \right)Z_{\mu}Z_{\nu} 
\left. +\left[\left(W_{\nu}^{+}\right)^2 + \left(W_{\nu}^{-}\right)^2\right](Z_{\mu})^2
\right\} -
$$
$$
-eg\cos\theta_W \biggl\{
W_\mu^{+}W_\mu^{-}A_{\nu}Z_{\nu} +
W_\nu^{+}W_\nu^{-}A_{\mu}Z_{\mu} -
$$
\begin{equation}
-\frac{1}{2}\left(W_\mu^{+}W_\nu^{-} + W_\nu^{+}W_\mu^{-} \right)\left(A_{\mu}Z_{\nu} + A_{\nu}Z_{\mu}\right)
\biggr\}.
\label{6g}
\end{equation}


The fermion sector is represented by  the lepton $L_L$ and
quark  $L_Q$ Lagrangians.
For first-generation particles, the lepton Lagrangian is written  in the form
\begin{equation}
L_{L,e}=L_l^{\dagger}i\tilde{\tau}_{\mu}D_{\mu}L_l + e_r^{\dagger}i\tau_{\mu}D_{\mu}e_r -
h_e[e_r^{\dagger}(\phi^{\dagger}L_l) +(L_l^{\dagger}\phi)e_r],
\label{eq14}
\end{equation}
where
$
L_l= \left(
\begin{array}{c}
	\nu_l\\
	e_{l}
\end{array} \right) \in \mathbf{C}_2
$
is the $SU(2)$ doublet,  $e_r $ is the $SU(2)$ singlet,
$\tau_{0}=\tilde{\tau_0}={\bf 1},$ $\tilde{\tau_k}=-\tau_k $, 
 $\; h_e$ is the Yukawa coupling,
 $e_r, e_l, \nu_l $ are the two-component Lorentz spinors. (We restrict ourselves to considering only particles. To take into account antiparticles, the fields must be four-component Dirac bispinors).
  Here $D_{\mu} $
are the covariant derivatives of the lepton fields
$$
D_\mu L_l=\partial_\mu L_l -i\frac{g}{\sqrt{2}}\left(W_{\mu}^{+}T_{+} + W_{\mu}^{-}T_{-} \right)L_l-
 i\frac{g}{\cos \theta_w}Z_\mu\left( T_3 -Q\sin^2 \theta_w  \right)L_l -ieA_\mu Q L_l,
$$
\begin{equation}
D_{\mu}e_r = \partial_\mu e_r -ig'QA_\mu e_r \cos \theta_w +ig'QZ_\mu e_r \sin \theta_w,
\label{eq5-10}
\end{equation}
where
$T_{\pm}=T_1\pm iT_2 $,
 $Q =Y+T_3$ is the generator of the $U(1)_{em}$ electromagnetic subgroup,
 $Y=\frac{1}{2}{\bf 1}$ is the hypercharge,
$ e=gg'(g^2+g'^2)^{-\frac{1}{2}} \;$ is the electron charge,and
$ \sin \theta_w=eg^{-1}.$


According to modern knowledge, all known leptons and quarks form three generations.
The next two lepton generations are introduced in a way similar  to (\ref{eq14}). They are left $SU(2)$-doublets
\begin{equation}
 \left(
\begin{array}{c}
	\nu_\mu\\
	\mu
\end{array} \right)_l, \quad
\left(
\begin{array}{c}
	\nu_\tau\\
	\tau
\end{array} \right)_l, \quad Y=-\frac{1}{2}
\label{eq14-1d}
\end{equation}
and right $SU(2)$-singlets:
$ 
\mu_r, \, \tau_r,\,  Y=-1.
$ 
In addition to $u$ and $d$ quarks of the first generation there is
$(c,s)$ and  $(t,b)$ quarks of the next generations, whose left fields
 \begin{equation}
 \left(
\begin{array}{c}
	c_l\\
	s_l
\end{array} \right), \quad
\left(
\begin{array}{c}
	t_l\\
	b_l
\end{array} \right), \quad Y=\frac{1}{6}
\label{eq14-1Q}
\end{equation}
are described by the  $SU(2)$-doublets and the right fields are  $SU(2)$-singlets:
$
c_r,\, t_r, \; Y=\frac{2}{3}; \;
s_r,\, b_r, \; Y=-\frac{1}{3}. \;
$
Their Lagrangians are introduced in a way similar  to (\ref{4}).
Full lepton and quark  Lagrangians are obtained by summing over all generations.
In what follows we will discuss only the first generations of leptons and quarks.

In terms of electron and neutrino fields the lepton  Lagrangian (\ref{eq14}) is written in the form
$$
L_L=e_l^{\dagger}i\tilde{\tau}_{\mu}\partial_{\mu}e_l +
e_r^{\dagger}i\tau_{\mu}\partial_{\mu}e_r
-m_e(e_r^{\dagger}e_l + e_l^{\dagger} e_r)+
$$
$$
+\frac{g\cos 2\theta_w}{2\cos \theta_w}e_l^{\dagger}\tilde{\tau}_{\mu}Z_{\mu}e_l
-ee_l^{\dagger}\tilde{\tau}_{\mu}A_{\mu}e_l 
g'\cos \theta_w e_r^{\dagger}\tau_{\mu}A_{\mu}e_r +
$$
$$
+  g'\sin \theta_w e_r^{\dagger}\tau_{\mu}Z_{\mu}e_r
+ \nu_l^{\dagger}i\tilde{\tau}_{\mu}\partial_{\mu}\nu_l
 +
 \frac{g}{2\cos \theta_w} \nu_l^{\dagger}\tilde{\tau}_{\mu}Z_{\mu}\nu_l +
$$
\begin{equation}
   +\frac{g}{\sqrt{2}}\left[ \nu_l^{\dagger}\tilde{\tau}_{\mu}W_{\mu}^{+}e_l +
 e_l^{\dagger}\tilde{\tau}_{\mu}W_{\mu}^{-}\nu_l\right].
\label{g15-4}
\end{equation}

The quark Lagrangian is constructed in a similar way
$$
L_Q=Q_l^{\dagger}i\tilde{\tau}_{\mu}D_{\mu}Q_l + 
u_r^{\dagger}i\tau_{\mu}D_{\mu}u_r +
d_r^{\dagger}i\tau_{\mu}D_{\mu}d_r -
$$
\begin{equation}
-h_d[d_r^{\dagger}(\phi^{\dagger}Q_l) +(Q_l^{\dagger}\phi)d_r]
-h_u[u_r^{\dagger}(\tilde{\phi}^{\dagger}Q_l) +(Q_l^{\dagger}\tilde{\phi})u_r],
\label{4}
\end{equation}
where the left quark fields make up the $SU(2)$ doublet
$
Q_l= \left(
\begin{array}{c}
	u_l\\
	d_{l}
\end{array} \right)\in \mathbf{C}_2,
$
the right fields   $u_r, d_r $ are $SU(2)$ singlets,
$\tilde{\phi}_i=\epsilon_{ik}\bar{\phi}_k, \epsilon_{00}=1, \epsilon_{ii}=-1$  make up a conjugate representation of the   $SU(2)$ group,
 $h_u, h_d$ are  the Yukawa couplings.
For particles all fields  $u_l, d_l, u_r, d_r $ are two-component
Lorentz spinors. The covariant derivatives of the quark
fields are
$$
D_{\mu}Q_l=\left(\partial_{\mu}- ig\sum_{k=1}^{3}\frac{\tau_k}{2}A^k_\mu -ig'\frac{1}{6}B_\mu  \right)Q_l,
$$
\begin{equation}
D_{\mu}u_r=\left(\partial_{\mu}-ig'\frac{2}{3}B_\mu  \right) u_r,\quad
D_{\mu}d_r=\left(\partial_{\mu}+ig'\frac{1}{3}B_\mu  \right) d_r.
\label{eq14-3Q}
\end{equation}

The quark  Lagrangian (\ref{4}) in terms of $u$ and $d$ quarks  fields can be written as
$$
L_Q=d_{l}^{\dagger}i\tilde{\tau}_{\mu}\partial_{\mu}d_{l}
+d_r^{\dagger}i\tau_{\mu}\partial_{\mu}d_r
-m_d(d_r^{\dagger}d_{l} + d_{l}^{\dagger}d_r)
-\frac{e}{3}d_{l}^{\dagger}\tilde{\tau}_{\mu}A_{\mu}d_{l} -
$$
$$
-\frac{g}{\cos \theta_w}\left(\frac{1}{2}-
\frac{2}{3}\sin^2\theta_w\right) d_{l}^{\dagger}\tilde{\tau}_{\mu}Z_{\mu}d_{l} 
 -\frac{1}{3}g'\cos\theta_w d_r^{\dagger}\tau_{\mu}A_{\mu}d_r +
$$
$$
+\frac{1}{3}g'\sin\theta_w d_r^{\dagger}\tau_{\mu}Z_{\mu}d_r 
+ u_{l}^{\dagger}i\tilde{\tau}_{\mu}\partial_{\mu}u_{l}
+u_r^{\dagger}i\tau_{\mu}\partial_{\mu}u_r
-m_u(u_r^{\dagger}u_{l} + u_{l}^{\dagger}u_r) +
$$
$$
+\frac{g}{\cos \theta_w}\left(\frac{1}{2}-\frac{2}{3}\sin^2\theta_w\right) u_{l}^{\dagger}\tilde{\tau}_{\mu}Z_{\mu}u_{l} 
+\frac{2e}{3}u_{l}^{\dagger}\tilde{\tau}_{\mu}A_{\mu}u_{l} +
$$
$$
+\frac{g}{\sqrt{2}}\left[ u_{l}^{\dagger}\tilde{\tau}_{\mu}W^{+}_{\mu}d_{l} +
d_{l}^{\dagger}\tilde{\tau}_{\mu}W^{-}_{\mu}u_{l}\right] 
+\frac{2}{3}g'\cos\theta_w u_r^{\dagger}\tau_{\mu}A_{\mu}u_r -
$$
\begin{equation}
-\frac{2}{3}g'\sin\theta_w u_r^{\dagger}\tau_{\mu}Z_{\mu}u_r,
\label{QL}
\end{equation}
where $m_e=h_ev/\sqrt{2}$ and $m_u=h_uv/\sqrt{2},\; m_d=h_dv/\sqrt{2} $ represent electron and quark masses.


\section{Electroweak Model at High Energies}

There are two ways to describe the action of a contracted group in a space.
The traditional way is to act by a matrix group with real or complex elements on vectors with similar components
$$
\left( \begin{array}{c}
	z'_1 \\
	z'_2
\end{array} \right)
=\left(\begin{array}{cc}
	\alpha & \beta   \\
-{\epsilon}^2\bar{\beta}	 & \bar{\alpha}
\end{array} \right)
\left( \begin{array}{c}
	z_1 \\
	z_2
\end{array} \right),
$$
\begin{equation}
\det u(\epsilon)=|\alpha|^2+{\epsilon^2}|\beta|^2=1, \quad u(\epsilon)u^{\dagger}(\epsilon)=1.
\label{51}
\end{equation}
In the limit  ${\epsilon} \rightarrow 0 $, this matrix has the form
$$
u(0)=\left(\begin{array}{cc}
	\alpha & \beta   \\
0	 & \bar{\alpha}
\end{array} \right), \quad
\alpha = e^{i\gamma},\;\; \gamma \in R
$$
and obviously  belongs to the group $E(2)$.

We will introduce a contracted group $SU(2;\epsilon)$ and corresponding space $\mathbf{C}_2(\epsilon)$
by the consistent rescaling  of the group $SU(2)$  and the  space ${\bf C}_2$ 
$$
\left( \begin{array}{c}
	z'_1 \\
{\epsilon}	z'_2
\end{array} \right)
=\left(\begin{array}{cc}
	\alpha & {\epsilon}\beta   \\
-{\epsilon}\bar{\beta}	 & \bar{\alpha}
\end{array} \right)
\left( \begin{array}{c}
	z_1 \\
{\epsilon}	z_2
\end{array} \right),
$$
\begin{equation}
\det u(\epsilon)=|\alpha|^2+{\epsilon^2}|\beta|^2=1, \quad u(\epsilon)u^{\dagger}(\epsilon)=1.
\label{5}
\end{equation}
%
Our approach  is based on the action of matrices with elements depending on the contraction parameter $\epsilon$ on vectors, the components of which also depend on this parameter.
The contraction parameter tending to zero is convenient for physical applications, but mathematically it can be equal to the nilpotent unit $\iota$, which has the property
$\iota \neq 0 $, but $\iota^2=0 $. Then the contracted matrix
$$
u(\iota)=\left(\begin{array}{cc}
	\alpha & \iota\beta   \\
-{\iota}\bar{\beta}		 & \bar{\alpha}
\end{array} \right), \quad
\alpha = e^{i\gamma},\;\; \gamma \in R.
$$
will not be diagonal, but will have nilpotent elements.
This approach   is detailed in [5] (formulae (1.17)--(1.24) and section 1.3.2, which describes the  $SU(2;\epsilon)$).

After  contraction the group  $SU(2;\epsilon=0)$  is isomorphic to Euclid group $E(2)$.
The contracted space ${\bf C}_2(\epsilon=0)$ is splitted  
on  the  base spanned by the $\left\{z_1\right\}$ coordinate and the  fiber spanned by the $\left\{z_2\right\}$  coordinate.
(The non-relativistic space--time is the best known example of a fiber space.
It has the one-dimensional base, which is interpreted as time,
 and three-dimensional fiber, which is interpreted as a prope space.)
 The unitary group $U(1)$ and its action in the space ${\bf C}_2(\epsilon=0)$ do not change and are given by (\ref{eq1qq}).

 The space ${\bf C}_2(\epsilon)$
 can be obtained from ${\bf C}_2$ by substitution of $z_2$ by $\epsilon z_2$, which  induces the other ones for Lie algebra generators
$T_1 \rightarrow \epsilon T_1,  T_2 \rightarrow \epsilon T_2, T_3 \rightarrow  T_3$. These new generators obey  the  comutation relations
\begin{equation}
[T_1,T_2]=i\epsilon^2T_3, \quad [T_3,T_1]=iT_2,\quad
 [T_2,T_3]=iT_1
\label{g8g}
\end{equation}
for Lie algebra $su(2;\epsilon).$
The structure of the algebra $su(2;\epsilon=0)$ is a semi-direct sum of  Abelian  subalgebra  $t_2=\{T_1,T_2\}$ and  subalgebra $u(1)=\{T_3\}:\;$
$
su(2;\epsilon)=t_2  \lplus  u(1).
$

 As far as the gauge fields take their values in Lie algebra, we can substitute the   gauge fields instead of transforming the generators, namely:
\begin{equation}
A_{\mu}^1 \rightarrow \epsilon A_{\mu}^1, \;\; A_{\mu}^2 \rightarrow \epsilon A_{\mu}^2, \;\;
A_{\mu}^3 \rightarrow A_{\mu}^3, \;\; B_{\mu} \rightarrow B_{\mu}.
\label{g15-bb}
\end{equation}
From the commutativity and associativity properties of multiplication by the scalar $\epsilon$, we have
$$
su(2;\epsilon)\ni \left\{   A_{\mu}^1(\epsilon T_1) + A_{\mu}^2(\epsilon T_2) + A_{\mu}^3T_3\right\}=
$$
\begin{equation}
= \left\{ (\epsilon A_{\mu}^1)T_1 + (\epsilon A_{\mu}^2)T_2 +
A_{\mu}^3T_3\right\}.
\label{g14-d}
\end{equation}
%
 The substitution  $\beta \rightarrow {\epsilon} \beta $ induces multiplication  of
the  standard  gauge fields (\ref{eq5-1})
\begin{equation}
W_{\mu}^{\pm} \rightarrow {\epsilon}W_{\mu}^{\pm}, \quad Z_{\mu} \rightarrow Z_{\mu},\quad A_{\mu} \rightarrow A_{\mu}.
\label{g15-b}
\end{equation}
The left lepton
$
L_l= \left(
\begin{array}{c}
	\nu_l\\
	e_{l}
\end{array} \right),\;
$
and quark
$
Q_l= \left(
\begin{array}{c}
	u_l\\
	d_{l}
\end{array} \right)
$
 fields are  $SU(2)$-doublets, so their  components are transformed in the similar way as the components of the vector $z$, namely:
\begin{equation}
 	 e_{l} \rightarrow \epsilon e_{l},  \quad  d_{l} \rightarrow \epsilon d_{l}, \quad
 	 \nu_l \rightarrow \nu_l, \quad  	u_l \rightarrow u_l.
\label{6}
\end{equation}
The right lepton and quark fields  are  $SU(2)$-singlets and therefore are not changed.

The next reason for inequality of the first and second doublet components is the special mechanism of spontaneous symmetry breaking, which is used to generate  mass of vector bosons and other elementary particles
of the model.
In this mechanism one of the Lagrangian $L_B$
ground states
$  \phi^{vac}=\left(\begin{array}{c}
	0  \\
	v
\end{array} \right), \;\;  A_\mu^k=B_\mu=0 \;$
 is taken as the model vacuum, and then low field excitations $ v+\chi(x) $ with respect to the second component of the vacuum vector are  considered.
Therefore, the Higgs boson field $ \chi $, constant $ v $,
and $ v $-dependent particle masses $m_p$ are multiplied
by the contraction parameter:
\begin{equation}
 	\chi  \rightarrow \epsilon \chi,  \quad  v \rightarrow \epsilon v, \quad 
 m_p \rightarrow \epsilon m_p, \quad p={\chi}, W, Z, e, u, d.
\label{7}
\end{equation}

As a result of  transformations (\ref{g15-b})--(\ref{7})
the  boson Lagrangian (\ref{eq2}), (\ref{eq3}) 
can be written in the form
\begin{equation}
 L_B(\epsilon)= - \frac{1}{4}{\cal Z}_{\mu\nu}^2 - \frac{1}{4}{\cal F}_{\mu\nu}^2 + \epsilon^2 L_{B,2}
 + \epsilon^3 gW_\mu^{+}W_\mu^{-}\chi  + \epsilon^4 L_{B,4},
\label{8}
\end{equation}
where
$$
 L_{B,4}= m_W^2W_\mu^{+}W_\mu^{-} -\frac{1}{2}m_{\chi}^2\chi^2 -\lambda v \chi^3  - \frac{\lambda}{4} \chi^4 +
$$
\begin{equation}
 +\frac{g^2}{4}\left(W_\mu^{+}W_\nu^{-} - W_\mu^{-}W_\nu^{+}\right)^2 +
\frac{g^2}{4}W_\mu^{+}W_\nu^{-}\chi^2,
\label{8-1}
\end{equation}
$$
L_{B,2}= \frac{1}{2}\left(\partial_\mu\chi \right)^2
 + \frac{1}{2}m_Z^2\left(Z_\mu\right)^2 -\frac{1}{2}{\cal W}_{\mu\nu}^{+}{\cal W}_{\mu\nu}^{-}+
$$
$$
 +\frac{g m_z}{2\cos \theta_W} \left(Z_{\mu}\right)^2 \chi +
 \frac{g^2 }{8\cos^2\theta_W} \left(Z_{\mu}\right)^2 \chi^2 -
$$
$$
-2ig\left(W_\mu^{+}W_\nu^{-} - W_\mu^{-}W_\nu^{+}\right)
\Bigl( {\cal F}_{\mu\nu}\sin \theta_W + {\cal Z}_{\mu\nu}\cos \theta_W \Bigr)  -
$$
$$
-\frac{i}{2}e \left[A_{\mu}\left({\cal W}_{\mu\nu}^{+}W_\nu^{-} - {\cal W}_{\mu\nu}^{-}W_\nu^{+}\right) +
 \frac{i}{2}e A_{\nu}\left({\cal W}_{\mu\nu}^{+}W_\mu^{-} - {\cal W}_{\mu\nu}^{-}W_\mu^{+}\right) \right] -
$$
$$
-\frac{i}{2}g\cos \theta_W  \left[Z_{\mu}\left({\cal W}_{\mu\nu}^{+}W_\nu^{-} - {\cal W}_{\mu\nu}^{-}W_\nu^{+}\right) - \right.
$$
$$
\left.  -Z_{\nu}\left({\cal W}_{\mu\nu}^{+}W_\mu^{-} - {\cal W}_{\mu\nu}^{-}W_\mu^{+}\right) \right] 
-\frac{e^2}{4} \left\{
\left[\left(W_{\mu}^{+}\right)^2 + \left(W_{\mu}^{-}\right)^2\right](A_{\nu})^2- \right.
$$
$$
-2\left(W_\mu^{+}W_\nu^{+} + W_\mu^{-}W_\nu^{-} \right)A_{\mu}A_{\nu} + 
 \left.
\left[\left(W_{\nu}^{+}\right)^2 + \left(W_{\nu}^{-}\right)^2\right](A_{\mu})^2
\right\} -
$$
$$
-\frac{g^2}{4}\cos\theta_W \left\{
\left[\left(W_{\mu}^{+}\right)^2 + \left(W_{\mu}^{-}\right)^2\right](Z_{\nu})^2 - \right.
$$
$$
\left. -2\left(W_\mu^{+}W_\nu^{+} + W_\mu^{-}W_\nu^{-} \right)Z_{\mu}Z_{\nu} +
\left[\left(W_{\nu}^{+}\right)^2 + \left(W_{\nu}^{-}\right)^2\right](Z_{\mu})^2
\right\} -
$$
$$
 -eg\cos\theta_W \biggl[
W_\mu^{+}W_\mu^{-}A_{\nu}Z_{\nu} +
W_\nu^{+}W_\nu^{-}A_{\mu}Z_{\mu} - 
$$
\begin{equation}
-\frac{1}{2}\left(W_\mu^{+}W_\nu^{-} + W_\nu^{+}W_\mu^{-} \right)\left(A_{\mu}Z_{\nu} + A_{\nu}Z_{\mu}\right)
\biggr].
\label{8-2}
\end{equation}

The lepton  Lagrangian (\ref{eq14}), (\ref{g15-4}) in terms of electron and neutrino fields takes the form
$$
L_{L}(\epsilon)= L_{L,0} + \epsilon^2 L_{L,2} =
$$
$$
=\nu_l^{\dagger}i\tilde{\tau}_{\mu}\partial_{\mu}\nu_l +
e_r^{\dagger}i\tau_{\mu}\partial_{\mu}e_r  +g'\sin \theta_w e_r^{\dagger}\tau_{\mu}Z_{\mu}e_r -
$$
$$
- g'\cos \theta_w e_r^{\dagger}\tau_{\mu}A_{\mu}e_r
 + \frac{g}{2\cos \theta_w} \nu_l^{\dagger}\tilde{\tau}_{\mu}Z_{\mu}\nu_l +
$$
$$
 +\epsilon^2\biggl\{ e_l^{\dagger}i\tilde{\tau}_{\mu}\partial_{\mu}e_l
-m_e(e_r^{\dagger}e_l + e_l^{\dagger} e_r)+
 \frac{g\cos 2\theta_w}{2\cos \theta_w}e_l^{\dagger}\tilde{\tau}_{\mu}Z_{\mu}e_l -
$$
\begin{equation}%
-ee_l^{\dagger}\tilde{\tau}_{\mu}A_{\mu}e_l 
+\frac{g}{\sqrt{2}}\left( \nu_l^{\dagger}\tilde{\tau}_{\mu}W_{\mu}^{+}e_l +
 e_l^{\dagger}\tilde{\tau}_{\mu}W_{\mu}^{-}\nu_l\right)\biggr\}.
\label{9}
\end{equation}

The quark  Lagrangian (\ref{4}), (\ref{QL})  
is written as
\begin{equation}
L_{Q}(\epsilon)= L_{Q,0} - \epsilon \, m_u(u_r^{\dagger}u_{l} + u_{l}^{\dagger}u_r) + \epsilon^2 L_{Q,2},
\label{12-1}
\end{equation}
where
$$
L_{Q,0}
=d_r^{\dagger}i\tau_{\mu}\partial_{\mu}d_r
+u_{l}^{\dagger}i\tilde{\tau}_{\mu}\partial_{\mu}u_{l}
+u_r^{\dagger}i\tau_{\mu}\partial_{\mu}u_r -
$$
$$
-\frac{1}{3}g'\cos\theta_w d_r^{\dagger}\tau_{\mu}A_{\mu}d_r
+ \frac{1}{3}g'\sin\theta_w d_r^{\dagger}\tau_{\mu}Z_{\mu}d_r
+\frac{2e}{3}u_{l}^{\dagger}\tilde{\tau}_{\mu}A_{\mu}u_{l} +
$$
$$
+\frac{g}{\cos \theta_w}\left(\frac{1}{2}-\frac{2}{3}\sin^2\theta_w\right) u_{l}^{\dagger}\tilde{\tau}_{\mu}Z_{\mu}u_{l} +
$$
\begin{equation}
+\frac{2}{3}g'\cos\theta_w u_r^{\dagger}\tau_{\mu}A_{\mu}u_r
-\frac{2}{3}g'\sin\theta_w u_r^{\dagger}\tau_{\mu}Z_{\mu}u_r,
\label{12-2}
\end{equation}
$$
L_{Q,2}=d_{l}^{\dagger}i\tilde{\tau}_{\mu}\partial_{\mu}d_{l}
- m_d(d_r^{\dagger}d_{l} + d_{l}^{\dagger}d_r)
-\frac{e}{3}d_{l}^{\dagger}\tilde{\tau}_{\mu}A_{\mu}d_{l} -
$$
$$
-\frac{g}{\cos \theta_w}\left(\frac{1}{2}-
\frac{2}{3}\sin^2\theta_w\right) d_{l}^{\dagger}\tilde{\tau}_{\mu}Z_{\mu}d_{l} +
$$
\begin{equation}
+\frac{g}{\sqrt{2}}\left[ u_{l}^{\dagger}\tilde{\tau}_{\mu}W^{+}_{\mu}d_{l} +
d_{l}^{\dagger}\tilde{\tau}_{\mu}W^{-}_{\mu}u_{l}\right].
\label{12}
\end{equation}

The complete  Lagrangian of the  Electroweak Model with contracted gauge group  is given by the sum
$
L_{EWM}(\epsilon)=L_B(\epsilon) + L_L(\epsilon) + L_Q(\epsilon)
$
and takes the form 
\begin{equation}
 L_{EWM}({\epsilon})= L_{0} + {\epsilon} L_{1}  
 +{\epsilon^2} L_{2} + {\epsilon^3} L_{3}  + {\epsilon^4} L_{4}.
\label{15}
\end{equation}
where
$$
L_{0}= L_{B,0}+L_{L,0}+L_{Q,0},\quad
 L_{1}=L_{Q,1}= - {\epsilon} \, m_u(u_r^{\dagger}u_{l} + u_{l}^{\dagger}u_r),
$$
\begin{equation}
L_{2}= L_{B,2}+L_{L,2}+L_{Q,2},\quad L_3=L_{B,3},\quad L_4=L_{B,4}.
\label{15-5}
\end{equation}

We assume that the contraction parameter is monotonous function   $\epsilon(T) $ of the temperature (or average  energy)   of the Universe
with the property
$\epsilon(T) \rightarrow 0 $ for $T \rightarrow \infty $.
When  ${\epsilon} \rightarrow 0 $, the terms with higher powers of  ${\epsilon} $ make
a smaller contribution as compared to the terms with lower  powers.
Thus, the Electroweak Model demonstrates
five stages of the behavior as ${\epsilon} \rightarrow 0 $, which differ by the powers of the contraction parameter.
This offers an opportunity for construction of intermediate limit models. 
One can take the Lagrangian $L_{0}$ for the initial limit system, then add
$L_1$ 
and obtain the second limit model with the Lagrangian
${\cal L}_1=L_{0} +  L_1 $. After that one can add
$ L_2 $ and obtain the third limit model
${\cal L}_2=L_{0} + L_1 +  L_2 $.
The last  limit model has the Lagrangian
${\cal L}_3=L_{0} +  L_1 +  L_2 +  L_3 $.

From the contraction of the Electroweak Model we can classify events in time as earlier-later, but we can not
determine their absolute time without additional assumptions.
At the level of classical gauge fields we can already  conclude that the $u$ quark  first restores  its mass in the evolution of the Universe.
Indeed the mass term of the $u$ quark in the Lagrangian (\ref{15})
$ L_1=-m_u(u_r^{\dagger}u_{l} + u_{l}^{\dagger}u_r) $
is proportional to the first power $\epsilon  $,
whereas the mass terms of $Z$ boson, electron and $d$ quark are multiplied by the second power of the contraction parameter
\begin{equation}
\epsilon^2\, \left[\frac{1}{2}m_Z^2\left(Z_\mu\right)^2 + m_e(e_r^{\dagger}e_l + e_l^{\dagger} e_r)+
 m_d(d_r^{\dagger}d_{l} + d_{l}^{\dagger}d_r)\right]. 
\label{40}
\end{equation}
 Massless charged $W$ bosons and Higgs boson $\chi$    appear at the same time. These particles  restore their masses after all other particles of the Electroweak Model because their mass terms are proportional to the fourth power $\epsilon^4 $.

The main part of the electroweak interactions  are  restored  in the epoch which corresponds to the second order of the contraction parameter. There is one term in Lagrangian (\ref{8})
$L_3=gW_\mu^+W_\mu^-\chi$  proportionate to $\epsilon^3$.
The final reconstruction of the electroweak interactions and the restoration of mass of all particles takes place at the last stage ($\approx\epsilon^4 $).

Two other generations of leptons (\ref{eq14-1d}) and quarks (\ref{eq14-1Q}) have similar properties:
for the "infinite" energy there are only massless right $\mu$ and $\tau$ muons, left $\mu$ and $\tau$ neutrinos, as well as massless left and right quarks $c_l, c_r, s_r, t_l, t_r, b_r$. $c$- and $t$ quarks first acquire their mass and after that $\mu$,  $\tau$ muons, $s$,  $b$ quarks  become massive.

\section{Quantum Chromodynamics}

Strong interactions of quarks are described by the Quantum Chromodynamics (QCD).
QCD is a gauge theory based on the local color degrees of freedom \cite{Em-2007}.
The QCD  gauge group is $SU(3)$, acting in three dimensional complex space ${\bf C}_3$ of  color quark states
$
q=
\left(\begin{array}{c}
 q_1\\
 q_2 \\
 q_3
 \end{array}
 \right) \equiv
\left(\begin{array}{c}
 q_R\\
 q_G \\
 q_B
 \end{array}
 \right)
 \in {\bf C}_3, \;
$
where $q(x)$ are quark fields $q=u, d, s, c, b, t $ and  $R$ (red), $G$ (green), $B$ (blue) are color degrees of freedom. The $SU(3)$ gauge bosons are called gluons. There are eight gluons in total, which
 are the force carrier  between quarks. The QCD Lagrangian is taken in the form
 \begin{equation}
{\cal L} =\sum_q \left( \bar{q}^i(i\gamma^\mu)(D_\mu)_{ij}q^j
  -m_q\bar{q}^iq_i \right)
-\frac{1}{4}\sum_{\alpha=1}^8
F_{\mu\nu}^\alpha F^{\mu\nu\, \alpha},
 \label{q1}
\end{equation}
where $D_\mu q$ are covariant derivatives of quark fields
\begin{equation}
D_\mu q =
\left(\partial_{\mu}-ig_s\left(\frac{\lambda^\alpha}{2}\right)A^\alpha_{\mu}\right)q,
 \label{q3}
\end{equation}
$g_s$ is  the strong coupling constant, $t^a=\lambda^a/2$ are the generators of $SU(3)$, $\lambda^a $ are Gell-Mann matrices
in the form
$$
\lambda^1=
\left(\begin{array}{ccc}
 0 & 1      & 0  \\
   1   & 0  & 0  \\
 0 & 0  & 0  \\
 \end{array}
 \right), \quad
\lambda^2=
\left(\begin{array}{ccc}
 0 & -i      & 0  \\
   i   & 0  & 0  \\
 0 & 0  & 0  \\
 \end{array}
 \right), \quad
 \lambda^3=
\left(\begin{array}{ccc}
    1   & 0  & 0  \\
 0  & -1     & 0  \\
 0  & 0  & 0  \\
 \end{array}
 \right), 
$$
$$
\lambda^4=
\left(\begin{array}{ccc}
 0 & 0  & 1  \\
 0 & 0  & 0  \\
   1   & 0  & 0  \\
 \end{array}
 \right), \quad
\lambda^5=
\left(\begin{array}{ccc}
 0 & 0  & -i     \\
 0 & 0  & 0  \\
  i    & 0  & 0  \\
 \end{array}
 \right), \quad
\lambda^6=
\left(\begin{array}{ccc}
 0 &  0  & 0  \\
   0   & 0  & 1   \\
 0 & 1  & 0  \\
 \end{array}
 \right), 
$$
  \begin{equation}
 \lambda^7=
\left(\begin{array}{ccc}
    0   & 0  & 0  \\
 0  & 0     & -i  \\
 0  & i  & 0  \\
 \end{array}
 \right), \quad
\lambda^8=\frac{1}{\sqrt{3}}
\left(\begin{array}{ccc}
 1 & 0  & 0  \\
 0 & 1  & 0  \\
   0   & 0  & -2  \\
 \end{array}
 \right).
 \label{q3-1}
\end{equation}
Gluon strength tensor is defined by the equation
\begin{equation}
F_{\mu\nu}^\alpha=\partial_{\mu} A_\nu^\alpha-\partial_{\nu} A_\mu^\alpha+
g_sf^{\alpha\beta\gamma}A_\mu^\beta A_\nu^\gamma.
 \label{q3-2}
\end{equation}
Here $ f^{\alpha\beta\gamma}  $ are the  structure constant of the algebra $su(3)$:
$[t^\alpha,t^\beta]=if^{\alpha\beta\gamma}t^\gamma,\; \alpha, \beta, \gamma=1,\ldots,8$.
They are antisymmetric on all indices and its nonzero values are as follows:
$$
f^{123}=1,\quad f^{147}=f^{246}=f^{257}=f^{345}=\frac{1}{2},
$$
\begin{equation}
f^{156}=f^{367}= -\frac{1}{2},\quad
f^{458}=f^{678}=\frac{\sqrt{3}}{2}.
 \label{q3-3}
\end{equation}

The choice of Gell-Mann matrices in the form (\ref{q3-1})  fix the basis in $SU(3)$. This enables us to write out  the covariant derivatives (\ref{q3}) in the explicit form
$$
D_\mu={\mathbf I}\partial_\mu -i\frac{g_s}{2}
\left(\begin{array}{cccccccc}
 A_\mu^3+\frac{1}{\sqrt{3}}A_\mu^8 & A_\mu^1-i A_\mu^2 & A_\mu^4-i A_\mu^5\\
 A_\mu^1+iA_\mu^2 & \frac{1}{\sqrt{3}}A_\mu^8- A_\mu^3 & A_\mu^6-i A_\mu^7\\
 A_\mu^4+iA_\mu^5 & A_\mu^6+i A_\mu^7 & -\frac{2}{\sqrt{3}} A_\mu^8 \\
 \end{array}
 \right)=
$$
 \begin{equation}
={\mathbf I}\partial_\mu
 -i\frac{g_s}{2}
\left(\begin{array}{cccccccc}
 A_\mu^{RR} & A_\mu^{RG} & A_\mu^{RB} \\
 A_\mu^{GR} & A_\mu^{GG} & A_\mu^{GB} \\
 A_\mu^{BR} & A_\mu^{BG} &  A_\mu^{BB} \\
 \end{array}
 \right),
 \label{q4}
\end{equation}
where
$$
A_\mu^{RR}=\frac{1}{\sqrt{3}}A_\mu^8+A_\mu^3, \quad A_\mu^{GG}= \frac{1}{\sqrt{3}}A_\mu^8- A_\mu^3,\quad
A_\mu^{BB}= -\frac{2}{\sqrt{3}} A_\mu^8,
$$
$$
A_\mu^{RR}+ A_\mu^{GG}+ A_\mu^{BB}=0, \quad
A_\mu^{GR}= A_\mu^1+iA_\mu^2 =\bar{A}_\mu^{RG},
$$
\begin{equation}
A_\mu^{BR}= A_\mu^4+iA_\mu^5 =\bar{A}_\mu^{RB}, \quad
A_\mu^{BG}= A_\mu^6+iA_\mu^7 =\bar{A}_\mu^{GB},
 \label{q4-1}
\end{equation}
and Lagrangian (\ref{q1})
$$
{\cal L} = \left( \bar{u}_i(i\gamma^\mu)(D_\mu)^{ij}u_j -m_u\bar{u}^iu_i  \right)
 + \ldots - \frac{1}{4}F_{\mu\nu}^\alpha F^{\mu\nu\, \alpha} \equiv
$$
 \begin{equation}
\equiv  L_u + \ldots - \frac{1}{4}F_{\mu\nu}^\alpha F^{\mu\nu\, \alpha},
 \label{q5}
\end{equation}
where only the $u$ quark part is given. Let us note, that in QCD  the special mechanism of spontaneous symmetry breaking is absent.  Therefore, gluons are massless particles.

The QCD Lagrangian has  rich dynamical content. It describes complicated hadron spectrum, color confinement, asymptotic freedom and many other effects.


\section{QCD at High Energies}

The contracted special unitary group $SU(3;\epsilon)$ is defined
by the action
\begin{equation}
q'({\epsilon})=\left(\begin{array}{c}
 q'_{1}\\
{\epsilon} q'_{2} \\
{\epsilon^2} q'_{3}
 \end{array}
 \right)=
\left(\begin{array}{ccc}
 u_{11}  &{\epsilon} u_{12} &{\epsilon^2} u_{13} \\
 {\epsilon} u_{21} & u_{22} & {\epsilon} u_{23} \\
 {\epsilon^2} u_{31} & {\epsilon} u_{32} & u_{33}
 \end{array}
 \right)
 \left(\begin{array}{c}
 q_{1}\\
{\epsilon} q_{2} \\
{\epsilon^2} q_{3}
 \end{array}
 \right)=U({\epsilon} )q({\epsilon} )
 \label{10}
\end{equation}
in the color space  $\mathbf{C}_3({\epsilon})$ as  ${\epsilon} \rightarrow 0 $.
Under this action the hermitian form
\begin{equation}
 q^{\dagger}(\epsilon )q(\epsilon )= \left|q_1\right|^2+  \epsilon^2\left|q_2\right|^2 +  \epsilon^4 \left|q_3\right|^2
 \label{q8}
\end{equation}
remains  invariant.
Transition from the classical group $SU(3)$ and space ${\bf C}_3$ to the group $SU(3;\epsilon)$ and space ${\bf C}_3(\epsilon)$ is given by the substitution
$$
q_1 \rightarrow q_1, \quad
q_2\rightarrow \epsilon  q_2,\quad q_3\rightarrow \epsilon^2  q_3,
$$
\begin{equation}
A_\mu^{GR}\rightarrow \epsilon A_\mu^{GR},\quad
A_\mu^{BG}\rightarrow \epsilon A_\mu^{BG},\quad
A_\mu^{BR}\rightarrow \epsilon^2 A_\mu^{BR},
 \label{q9}
\end{equation}
and diagonal gauge fields
$A_\mu^{RR}, A_\mu^{GG}, A_\mu^{BB} $
remain unchanged.
Substitutions (\ref{7}) and (\ref{q9}) lead to the quark part of the QCD Lagrangian in the form

\begin{equation}
{\cal L}_q(\epsilon )= L_q^{0}+ \epsilon L_q^{(1)} + \epsilon^2 L_q^{(2)} + \epsilon^3 L_q^{(3)}+ \epsilon^4 L_q^{(4)}+ \epsilon^5 L_q^{(5)},
 \label{q11-C}
\end{equation}
where
\begin{equation}
L_q^{0} =
\sum_q \Biggl\{ i\bar{q}_1\gamma^\mu\partial_{\mu} q_1 +
\frac{g_s}{2} \left|q_1\right|^2 \gamma^\mu \left(\frac{1}{\sqrt{3}}A_\mu^8+A_\mu^3\right)
\Biggr\},
 \label{Q-0}
\end{equation}
\begin{equation}
L_q^{(1)}= - \sum_q m_q\left|q_1\right|^2, \quad L_q^{(3)}= - \sum_q m_q\left|q_2\right|^2, \quad L_q^{(5)}= - \sum_q m_q\left|q_3\right|^2,
 \label{Q-1-3}
\end{equation}
$$
L_q^{(2)}= \sum_q \Biggl\{
i\bar{q}_2\gamma^\mu \partial_{\mu} q_2  +
 \frac{g_s}{2}\biggl(\left|q_2\right|^2 \gamma^\mu  \left(\frac{1}{\sqrt{3}}A_\mu^8-A_\mu^3\right)+
$$
\begin{equation}
+q_1\bar{q}_2\gamma^\mu  \left(A_\mu^1+iA_\mu^2\right)
+ \bar{q}_1q_2\gamma^\mu  \left(A_\mu^1-iA_\mu^2\right) \biggr)
\Biggr\},
 \label{Q-2}
\end{equation}
$$
L_q^{(4)}= \sum_q \Biggl\{
i\bar{q}_3\gamma^\mu \partial_{\mu} q_3  +
 \frac{g_s}{2}\biggl(-\frac{2}{\sqrt{3}} A_\mu^8 \left|q_3\right|^2 \gamma^\mu A_\mu^{BB}+
$$
$$
+q_1\bar{q}_3\gamma^\mu  \left(A_\mu^4+iA_\mu^5\right) + \bar{q}_1q_3\gamma^\mu  \left(A_\mu^4-iA_\mu^5\right)+
$$
\begin{equation}
+ q_2\bar{q}_3\gamma^\mu \left(A_\mu^6 +iA_\mu^7\right)+ \bar{q}_2q_3\gamma^\mu \left(A_\mu^6-iA_\mu^7\right) \biggr)\Biggr\}.
 \label{Q-4}
\end{equation}
The gluon part $L_{gl}=-\frac{1}{4}F_{\mu\nu}^\alpha F^{\mu\nu\, \alpha}$ of the Lagrangian has the form
\begin{equation}
{\cal L}_{gl}(\epsilon)=
L_{gl}^{0} + \epsilon^2 L_{gl}^{(2)} + \epsilon^4 L_{gl}^{(4)}+ \epsilon^6 L_{gl}^{(6)}+ \epsilon^8 L_{gl}^{(8)},
 \label{q11-D}
\end{equation}
where
 \begin{equation}
L_{gl}^{0}= -\frac{1}{4}\biggl\{  \left(\partial_{\mu} A_\nu^3-\partial_{\nu} A_\mu^3\right)^2 +
\left(\partial_{\mu} A_\nu^8-\partial_{\nu} A_\mu^8\right)^2  \biggr\},
 \label{GL-0}
\end{equation}
$$
L_{gl}^{(2)}= -\frac{1}{4}\Biggl\{
\biggl( \partial_{\mu} A_\nu^1-\partial_{\nu} A_\mu^1  +
 g_s \left(A_\mu^2 A_\nu^3-A_\mu^3 A_\nu^2 \right) \biggr)^2 +
$$
$$
+ \biggl(\partial_{\mu} A_\nu^6-\partial_{\nu} A_\mu^6 +
\frac{g_s}{2}\left[\left(A_\mu^3 A_\nu^7-A_\mu^7 A_\nu^3 \right)+\right.
$$
$$
 \left.  +\sqrt{3}\left(A_\mu^7 A_\nu^8-A_\mu^8 A_\nu^7  \right) \right]\biggr)^2 +
$$
$$
+ \biggl(\partial_{\mu} A_\nu^2-\partial_{\nu} A_\mu^2  -g_s\left(A_\mu^1 A_\nu^3-A_\mu^3 A_\nu^1 \right)  \biggr)^2 +
$$
$$
+\biggl(\partial_{\mu} A_\nu^7-\partial_{\nu} A_\mu^7  -\frac{g_s}{2}\left[A_\mu^3 A_\nu^6-A_\mu^6 A_\nu^3+ \right.
$$
$$
 \left. +\sqrt{3}\left(A_\mu^6 A_\nu^8-A_\mu^8 A_\nu^6 \right)   \right]\biggr)^2 +
$$
$$
 +g_s\biggl[\Bigl(2\left(A_\mu^1 A_\nu^2-A_\mu^2 A_\nu^1 \right) - A_\mu^6 A_\nu^7+A_\mu^7 A_\nu^6 \Bigr)
\left(\partial_{\mu} A_\nu^3-\partial_{\nu} A_\mu^3\right)  +
 $$
 \begin{equation}
 +\sqrt{3}\left(A_\mu^6 A_\nu^7-A_\mu^7 A_\nu^6 \right) \left(\partial_{\mu} A_\nu^8-\partial_{\nu} A_\mu^8\right)   \biggr]  \Biggr\},
 \label{GL-2}
\end{equation}
$$
L_{gl}^{(4)}
= -\frac{1}{4}\Biggl\{
\left(\partial_{\mu} A_\nu^4-\partial_{\nu} A_\mu^4  \right)^2+
\left(\partial_{\mu} A_\nu^5-\partial_{\nu} A_\mu^5  \right)^2 +
$$
$$
+g_s\biggl[ \Bigl(A_\mu^4 A_\nu^7-A_\mu^7 A_\nu^4-  A_\mu^5 A_\nu^6+A_\mu^6 A_\nu^5  \Bigr)
\left(\partial_{\mu} A_\nu^1-\partial_{\nu} A_\mu^1\right)  +
$$
$$
+\Bigl(A_\mu^4 A_\nu^6-A_\mu^6 A_\nu^4+A_\mu^5 A_\nu^7-A_\mu^7 A_\nu^5\Bigr) \left(\partial_{\mu} A_\nu^2-\partial_{\nu} A_\mu^2\right)  -
$$
$$
-\Bigl(A_\mu^1 A_\nu^7-A_\mu^7 A_\nu^1+A_\mu^2 A_\nu^6-A_\mu^6 A_\nu^2
+A_\mu^3 A_\nu^5-A_\mu^5 A_\nu^3 -
$$
$$
-\sqrt{3} \left( A_\mu^5 A_\nu^8-A_\mu^8 A_\nu^5 \right)\Bigr) \left(\partial_{\mu} A_\nu^4-\partial_{\nu} A_\mu^4\right) +
$$
$$
+\Bigl(A_\mu^1 A_\nu^6-A_\mu^6 A_\nu^1 - A_\mu^2 A_\nu^7+A_\mu^7 A_\nu^2+
A_\mu^3 A_\nu^4-A_\mu^4 A_\nu^3-
$$
$$
-\sqrt{3}\left(A_\mu^4 A_\nu^8-A_\mu^8 A_\nu^4 \right)\Bigr) \left(\partial_{\mu} A_\nu^5-\partial_{\nu} A_\mu^5\right) +
$$
$$
+\Bigl(A_\mu^2 A_\nu^4-A_\mu^4 A_\nu^2-A_\mu^1 A_\nu^5+A_\mu^5 A_\nu^1 \Bigr) \left(\partial_{\mu} A_\nu^6-\partial_{\nu} A_\mu^6\right)  +
$$
$$
+\Bigl(A_\mu^1 A_\nu^4-A_\mu^4 A_\nu^1+A_\mu^2 A_\nu^5-A_\mu^5 A_\nu^2\Bigr) \left(\partial_{\mu} A_\nu^7-\partial_{\nu} A_\mu^7\right)  +
$$
$$
+\sqrt{3} \left(A_\mu^4 A_\nu^5-A_\mu^5 A_\nu^4\right) \left(\partial_{\mu} A_\nu^8-\partial_{\nu} A_\mu^8\right)  \biggr] +
$$
$$
+g_s^2\biggl[ \left(A_\mu^1 A_\nu^2-A_\mu^2 A_\nu^1\right)^2+ \left(A_\mu^6 A_\nu^7-A_\mu^7 A_\nu^6\right)^2 -
$$
$$
 - \left(A_\mu^1 A_\nu^2-A_\mu^2 A_\nu^1\right)\left(A_\mu^6 A_\nu^7-A_\mu^7 A_\nu^6\right) -
$$
$$
- \left(A_\mu^1 A_\nu^3-A_\mu^3 A_\nu^1\right)\Bigl(A_\mu^4 A_\nu^6-A_\mu^6 A_\nu^4 +
A_\mu^5 A_\nu^7-A_\mu^7 A_\nu^5\Bigr) +
$$
$$
+\left(A_\mu^2 A_\nu^3-A_\mu^3 A_\nu^2\right)\Bigl(A_\mu^4 A_\nu^7-A_\mu^7 A_\nu^4
-A_\mu^5 A_\nu^6+A_\mu^6 A_\nu^5\Bigr) +
$$
$$
+\frac{1}{2}\Bigl(A_\mu^3 A_\nu^7-A_\mu^7 A_\nu^3+ \sqrt{3}\left(A_\mu^7 A_\nu^8-A_\mu^8 A_\nu^7\right)\Bigr)\times
$$
$$
\times\Bigl(A_\mu^2 A_\nu^4-A_\mu^4 A_\nu^2 -A_\mu^1 A_\nu^5+A_\mu^5 A_\nu^1\Bigr)-
$$
$$
 -\frac{1}{2}\Bigl(A_\mu^3 A_\nu^6-A_\mu^6 A_\nu^3+ \sqrt{3}\left(A_\mu^6 A_\nu^8-A_\mu^8 A_\nu^6\right)\Bigr)\times
$$
$$
\times \Bigl(A_\mu^1 A_\nu^4-A_\mu^4 A_\nu^1 +A_\mu^2 A_\nu^5-A_\mu^5 A_\nu^2\Bigr)+
$$
$$
+\frac{1}{2}\Bigl(A_\mu^1 A_\nu^7-A_\mu^7 A_\nu^1 +A_\mu^2 A_\nu^6-A_\mu^6 A_\nu^2+
$$
$$
 +A_\mu^3 A_\nu^5-A_\mu^5 A_\nu^3 -\sqrt{3}\left(A_\mu^5 A_\nu^8-A_\mu^8 A_\nu^5\right)\Bigr)^2 +
$$
$$
+\frac{1}{2}\Bigl(A_\mu^1 A_\nu^6-A_\mu^6 A_\nu^1 -A_\mu^2 A_\nu^7+A_\mu^7 A_\nu^2+
$$
\begin{equation}
 +A_\mu^3 A_\nu^4-A_\mu^4 A_\nu^3 -\sqrt{3}\left(A_\mu^4 A_\nu^8-A_\mu^8 A_\nu^4\right)\Bigr)^2 \biggr] \Biggr\},
\label{GL-4}
\end{equation}
$$
L_{gl}^{(6)}
= -\frac{g_s^2}{16}\Biggl\{\Bigl(A_\mu^4 A_\nu^7-A_\mu^7 A_\nu^4 -
A_\mu^5 A_\nu^6+A_\mu^6 A_\nu^5  \Bigr)^2 +
$$
$$
\Bigl(A_\mu^4 A_\nu^6-A_\mu^6 A_\nu^4+
A_\mu^5 A_\nu^7-A_\mu^7 A_\nu^5 \Bigr)^2 +
$$
$$
+ \Bigl(A_\mu^2 A_\nu^4-A_\mu^4 A_\nu^2-A_\mu^1 A_\nu^5+A_\mu^5 A_\nu^1  \Bigr)^2 +
$$
$$
+ \Bigl(A_\mu^1 A_\nu^4-A_\mu^4 A_\nu^1+
A_\mu^2 A_\nu^5-A_\mu^5 A_\nu^2 \Bigr)^2 +
$$
\begin{equation}
 +4\Bigl( A_\mu^1 A_\nu^2-A_\mu^2 A_\nu^1+
A_\mu^6 A_\nu^7-A_\mu^7 A_\nu^6 \Bigr)
\left(A_\mu^4 A_\nu^5-A_\mu^5 A_\nu^4 \right) \Biggr\},
\label{GL-6}
\end{equation}
\begin{equation}
L_{gl}^{(8)}=
-\frac{g_s^2}{4}\left(A_\mu^4 A_\nu^5-A_\mu^5 A_\nu^4 \right)^2.
 \label{GL-8}
\end{equation}

Thus, the Lagrangian of the modified QCD can be represented as an expansion in powers of the contraction parameter
\begin{equation}
{\cal L}_{QCD}(\epsilon)={\cal L}^{0} +\epsilon {\cal L}^{(1)} +\epsilon^2 {\cal L}^{(2)} +\epsilon^3 {\cal L}^{(3)} +\epsilon^4 {\cal L}^{(4)}
 +\epsilon^5 {\cal L}^{(5)} +\epsilon^6 {\cal L}^{(6)} +\epsilon^8 {\cal L}^{(8)},
 \label{QCD-1}
\end{equation}
where
$$
{\cal L}^{0}= L_q^{0} + L_{gl}^{0}, \quad {\cal L}^{(1)}= L_q^{(1)}, \quad {\cal L}^{(2)}= L_q^{(2)} + L_{gl}^{(2)}, \quad
{\cal L}^{(3)}= L_q^{(3)},
$$
\begin{equation}
{\cal L}^{(4)}= L_q^{(4)} + L_{gl}^{(4)}, \quad {\cal L}^{(5)}= L_q^{(5)}, \quad {\cal L}^{(6)}= L_{gl}^{(6)}, \quad {\cal L}^{(8)}= L_{gl}^{(8)}.
 \label{QCD-2}
\end{equation}
According to our  hypothesis, the contraction parameter is a monotonic function of temperature
$\epsilon \rightarrow 0 $ as $ T \rightarrow \infty $.
Very high ("ifinite") temperatures can exist  at the first stages of the Big Bang immediately after inflation  in the pre-electroweak epoch \cite{GoR-11,L-1990}.

\section{Estimation  of boundary values in the evolution of the Universe}

As noted,  the contraction of the gauge group of  the Standard Model makes it possible  to  order   in time different stages of its development,
but does not allow determining their absolute dates.
 For this additional assumptions are needed.
We assume that the contraction parameters for QCD and the Electroweak  Model  are the same.

Further, we assume that the Electroweak Model  is fully restored at its characteristic  temperature $E_4=100\, $ GeV,
and the complete reconstruction of the QCD occurs at the temperature $E_8=0,2\, $ GeV.
Let  $\Delta$ be  the cutoff  level for
$\epsilon^k, \; k=1,2,4,5,6,8$, i.e. when $\epsilon^k < \Delta$ all Lagrangians terms  proportional to $\epsilon^k$  are negligibly small.
Also we will assume that the contraction parameter is proportional to the inverse  temperature
\begin{equation}
\epsilon(T)=\frac{A}{T},
 \label{q14}
\end{equation}
where $ A=const. $
From the QCD equation
${\epsilon^8}(T_8)=A^8T_8^{-8}= \Delta $ we obtain
 $A=T_8\Delta^{1/8}=0,2\Delta^{1/8}$ GeV.
 Using the equation for the  $k$th pover
${\epsilon^k}(T_k)=A^kT_k^{-k}=\Delta, $
we have
\begin{equation}
T_k= 
T_8\Delta^{\frac{k-8}{8k}}\approx 10^{\frac{88-15k}{4k}}\, \mbox{GeV}
 \label{q15}
\end{equation}
 and after simple calculations one easily obtains the  boundary values (in GeV)
\begin{equation}
T_1=10^{18},\;\; T_2=10^7,\;\; T_3=10^3,\;\; T_4=10^2,\;\; T_5=4,\;\; T_6=1,\;\; T_8=2\cdot10^{-1}.
 \label{14}
\end{equation}
The  estimate of the "infinity" energy
$T_1\approx 10^{18}\,$ GeV is comparable with the Planck energy $\approx 10^{19}\,$  GeV,
at which it is necessary to take into account the gravitational effects.
Thus, the resulting evolution of elementary particles does not go beyond the problems described by  electroweak and strong interactions.

%

\section{Evolution of Particles}

Combining the modified Lagrangians of the electroweak model (\ref{15}) and quantum chromodynamics (\ref{QCD-1}), we arrive at the Lagrangian of the Standard Model, represented as an expansion in powers of the contraction parameter
\begin{equation}
{\cal L}_{SM}(\epsilon)= {\cal L}_{0} +\epsilon {\cal L}_{1} +\epsilon^2 {\cal L}_{2} +\epsilon^3 {\cal L}_{3} +\epsilon^4 {\cal L}_{4}
 +\epsilon^5 {\cal L}_{5} +\epsilon^6 {\cal L}_{6} +\epsilon^8 {\cal L}_{8},
 \label{SM-1}
\end{equation}
where
$$
{\cal L}_{0}= L_{0} + L_q^{0} + L_{gl}^{0}, \quad
{\cal L}_{1}=   L_1+{\cal L}^{(1)}=
L_{Q,1} +L_q^{(1)}, \quad
{\cal L}_{2}= L_2 + L_q^{(2)} + L_{gl}^{(2)},
$$
\begin{equation}
{\cal L}_{3}= L_q^{(3)},\quad
{\cal L}_{4}= L_4 + L_q^{(4)} + L_{gl}^{(4)}, \quad
{\cal L}_{5}= L_q^{(5)}, \quad {\cal L}_{6}= L_{gl}^{(6)}, \quad {\cal L}_{8}= L_{gl}^{(8)}.
 \label{SM-2}
\end{equation}
The properties of particles change from epoch to epoch and are described by  terms in the corresponding Lagrangians.
We may draw some conclusions  about the properties of particles and their interactions in different  epochs
even at the level of the classical fields.

In the "infinite" temperature   limit ($\epsilon =0$, $ T> 10^{18}$ GeV) Lagrangian of the Electroweak Model (\ref{15}) is written as
$$
L_{0}= - \frac{1}{4}{\cal Z}_{\mu\nu}^2 - \frac{1}{4}{\cal F}_{\mu\nu}^2 +
\nu_l^{\dagger}i\tilde{\tau}_{\mu}\partial_{\mu}\nu_l
+u_{l}^{\dagger}i\tilde{\tau}_{\mu}\partial_{\mu}u_{l}+
$$
\begin{equation}
+e_r^{\dagger}i\tau_{\mu}\partial_{\mu}e_r +
d_r^{\dagger}i\tau_{\mu}\partial_{\mu}d_r
+u_r^{\dagger}i\tau_{\mu}\partial_{\mu}u_r + L_{0}^{int}(A_{\mu},Z_{\mu}),
\label{15-dop}
\end{equation}
i.e., the limit model contains only  {\it massless particles}:
photons $A_{\mu}$ and neutral  bosons $Z_{\mu}$, left quarks $u_{l}$ and neutrinos $\nu_l$, right electrons $e_r $ and  quarks $ u_r, d_r $.
 This phenomenon has a simple physical explanation: the temperature is so high, that the particle mass is  a negligibly small quantity in
  comparison  to the kinetic energy.
 The electroweak interactions become long-range ones, since  they are transferred  by  massless  $Z$ bosons and photons.
the charged fields corresponding to the translation subgroup are not included in the limit Lagrangians
The charged   boson fields $W^{\pm}_{\mu}$ corresponding  to the translation subgroup   
are not included  in the limit Lagrangians  (\ref{15-dop}) and  (\ref{14kk}).

We see from  the interaction Lagrangian
$$
L_{0}^{int}(A_{\mu},Z_{\mu})=
$$
$$
=\frac{g}{2\cos \theta_w} \nu_l^{\dagger}\tilde{\tau}_{\mu}Z_{\mu}\nu_l
+\frac{2e}{3}u_{l}^{\dagger}\tilde{\tau}_{\mu}A_{\mu}u_{l}
+g'\sin \theta_w e_r^{\dagger}\tau_{\mu}Z_{\mu}e_r +
$$
$$
+\frac{g}{\cos \theta_w}\left(\frac{1}{2}-\frac{2}{3}\sin^2\theta_w\right) u_{l}^{\dagger}\tilde{\tau}_{\mu}Z_{\mu}u_{l}
 - g'\cos \theta_w e_r^{\dagger}\tau_{\mu}A_{\mu}e_r-
$$
$$
-\frac{1}{3}g'\cos\theta_w d_r^{\dagger}\tau_{\mu}A_{\mu}d_r + \frac{1}{3}g'\sin\theta_w d_r^{\dagger}\tau_{\mu}Z_{\mu}d_r +
$$
\begin{equation}
 + \frac{2}{3}g'\cos\theta_w u_r^{\dagger}\tau_{\mu}A_{\mu}u_r
-\frac{2}{3}g'\sin\theta_w u_r^{\dagger}\tau_{\mu}Z_{\mu}u_r.
\label{14kk}
\end{equation}
that particles of different kind do not interact with one another. For example,  neutrinos  interact only with each other through   neutral currents.
 All other particles are charged and interact with particles of the same sort by massless $Z_{\mu}$ bosons and photons.
 Particles of different kind do not interact.
This looks like a sort of stratification of the Electroweak Model with  particles of the same kind in each layer.

In the "infinite" temperature limit
 only two components of the gluon strength tensor are nonzero
$$
F_{\mu\nu}^3=\partial_{\mu} A_\nu^3-\partial_{\nu} A_\mu^3=\frac{1}{2}\left(F_{\mu\nu}^{RR}-F_{\mu\nu}^{GG}\right),
$$
\begin{equation}
F_{\mu\nu}^8=\partial_{\mu} A_\nu^8-\partial_{\nu} A_\mu^8=\frac{\sqrt{3}}{2}\left(F_{\mu\nu}^{RR}+F_{\mu\nu}^{GG}\right),
 \label{q12}
\end{equation}
so we can  write out the limiting QCD Lagrangian  explicitly
$$
{\cal L}^{0}=L_q^{0}+L_{gl}^{0}=
\sum_q \biggl\{ i\bar{q}_R\gamma^\mu\partial_{\mu} q_R
+\frac{g_s}{2} \left|q_R\right|^2 \gamma^\mu A_\mu^{RR} \biggr\} -
$$
 \begin{equation}
-\frac{1}{4}\left(F_{\mu\nu}^{RR}\right)^2
-\frac{1}{4}\left(F_{\mu\nu}^{GG}\right)^2
-\frac{1}{4}F_{\mu\nu}^{RR}F_{\mu\nu}^{GG}.
 \label{q13}
\end{equation}
 From ${\cal L}_{0}$ we
 conclude that in this limit   
 only  dynamic terms for one  color component of massless quarks
survive, i.e.,  quarks become  monochromatic.
Terms  describing  interaction of these components with $R$ gluons also persist.
Besides $R$ gluons there are also $G$ gluons, which do not interact with the quarks.
Thus, there is stratification in the QCD sector as well.

The limit Lagrangian ${\cal L}_{0}$ can be considered as a good approximation after  the  Big Bang,
just as the nonrelativistic   mechanics is a good  approximation of the relativistic one at low velocities.

In the temperature interval  $  10^{18}$ GeV $\geq  T > 10^7$ GeV the terms
\begin{equation}
{\cal L}_{1}=
- \sum_q  m_q\left|q_1\right|^2 - \sum_{\tilde q} m_{\tilde q}({\tilde q}_r^{\dagger}{\tilde q}_{l} + {\tilde q}_{l}^{\dagger}{\tilde q}_r), \quad
q=u,d,c,s,t,b, \;\; {\tilde q}=u,c,t
 \label{C1}
\end{equation}
  are added to the Lagrangian $ {\cal L}_{0}$.
The mass terms of 
quarks appear in the Lagrangian.  Therefore, all quarks restore their masses in the process of  evolution of the Universe in a given interval. 

In the  interval  $  10^{7}$ GeV $\geq  T > 10^3$ GeV  the electron  mass terms  
$
{{\epsilon^2}}\, \left[m_e(e_r^{\dagger}e_l + e_l^{\dagger} e_r)\right]
$
appear in the Lagrangian
The same is true for the $\mu$ and $\tau$ leptons. All these particles become massive at this epoch.

Quarks acquire the second color degree of freedom and
 scalar Higgs boson began to interact with charged $W$ bosons
 between $  10^{3}$ GeV  and $10^2$ GeV due to Lagrangian
$
{\cal L}_{3}= - \sum_q  m_q\left|q_2\right|^2 + gW_{\mu}^{+}W_{\mu}^{-}\chi.
$
The main part of the electroweak and color interactions is restored   in this two last stages. 

In the next interval $  10^{2}$ GeV $\geq T> 4 $ GeV  charged $W$ bosons and the Higgs boson $\chi$ are the last to restore mass.
More complex interactions arise, such as self-interactions of the Higgs boson, interactions of two Higgs bosons with two charged bosons, and four charged bosons.
The Electroweak Model is ultimately restored.
Quarks acquire the third  color degree of freedom between $4$ GeV  and $1$ GeV due to the terms
${\cal L}_{5}= - \sum_q  m_q\left|q_3\right|^2. $

At energies  1 Gev $\geq T >$ 0.2 GeV there exist all color interactions except
$L_{gl}^{(8)}= -\frac{g_s^2}{4} \Bigl(A_{\mu}^4A_{\nu}^5 - A_{\mu}^5A_{\nu}^4\Bigr)^2. $
Finally, at $ T \leq $ 0.2 GeV the Standard Model is entirely restored.

\section{Conclusions}

We have investigated   the high-temperature  limit of the Standard Model  which was obtained from the  first principles of the gauge theory as the contraction of its gauge group.
It was shown that if the mathematical contraction parameter is taken to be inversely proportional   to  the temperature of the Universe,
then  its zero limit corresponds to the "infinite" temperature of the order of the Planck energy of $10^{19}$ GeV.
In the process of the evolution of the Universe,
the Standard Model passes  through several stages, which are distinguished by the powers of the contraction parameter,
what gives the opportunity to classify them in time as earlier-later. To determine the absolute dates of these  stages the additional assumptions were used, namely: the inverse temperature dependence of contraction parameter $\epsilon $  and  the cutoff  level $\Delta $ for $\epsilon^k $. The unknown parameters are determined with consideration of typical QCD and Electroweak Model energies.

The exact expressions for the respective Lagrangians for any stage in the Standard Model evolution are obtained.
On the base of decompositions (\ref{15}), (\ref{QCD-1}), and (\ref{SM-1}),
 the intermediate Lagrangians $ {\cal L}_{k} $ for any temperature scale  are constructed.
It gives an opportunity to draw  conclusions on the interactions and properties of the elementary particles in each of the considered epochs.

The evolution of the elementary particles and their interactions in the early Universe obtained with the help of the contractions of the gauge groups of the Standard Model does not contradict  the canonical one  \cite{Em-2007},
according to which the QCD phase transitions take place later then the electroweak phase transitions. The developed evolution of the Standard Model present the basis for a more detailed analysis of different phases in the formation of leptons and quark-gluon plasma,
in view of the fact that
 the terms $L_{gl}^{(6)} $ and $L_{gl}^{(8)} $ 
 in the gluon Lagrangian (\ref{q11-D}) 
 become negligible small at temperatures from $0.2$  GeV to $100$ GeV and in the temperature interval of $100$ GeV to
$1000$ GeV only the interaction of the Higgs boson with
charged $W$ bosons is restored.


\end{document}